\documentstyle[manuscript, osa, osa10]{revtex}
\input epsf
\title{Surface and internal scattering in microspheres: limits imposed on the
Q-factor and mode splitting}

\author{M.L.Gorodetsky, V.S.Ilchenko, A.D.Pryamikov}
\address{M.L.Gorodetsky, A.D.Pryamikov -- Faculty of Physics, 
Moscow State University, 119899, Moscow, Russia, V.S.Ilchenko --
Jet Propulsion Laboratory, California Institute of Technology, 4800,
Oak Grove Dr, Pasadena, CA 91109-8099}

\begin{document}

\maketitle

\begin{abstract}
 Accurate calculation of internal and surface scattering losses in fused
 silica microspheres is done. We show that in microspheres internal
 scattering is partly inhibited as compared to losses in the bulk material.
 We pay attention on the effect of frozen thermodynamical capillary waves
 on surface roughness. We calculate also the value of mode splitting due 
 to backscattering and other effects of this backscattering.
\end{abstract}

Optical microsphere resonators \cite{first} working on whispering gallery modes
combine several unique features -- very high quality factor, small size and
effective volume of field localization with low cost, that make them very
attractive for future applications in optoelectronics and measurement science.
Such microresonators may be for example used as interferometers and filters
with record finesse \cite{filter,Kimbles}, for QED experiments
\cite{first,Kimbleq,haroche} and for diode laser stabilization \cite{laser}.
The last application as for now looks the most intriguing as it allows to
create centimeter size cheap tunable laser with kilohertz linewidth.

Microspheres are prepared from fused silica or other pure glassy medium
by autoforming under the action of surface tension in the flame of a
microburner or in the $CO_2$ beam.

The properties of the microspheres are analyzed theoretically and experimentally
rather wide. The theory of whispering gallery modes is well known from
electrodynamics \cite{Stratton} and methods of optimal coupling with them
were elaborated and experimentally confirmed \cite{coupl}. In \cite{nonlin}
nonlinear properties of fused silica microspheres were investigated
theoretically and experimentally. Mechanisms limiting the quality factor of
microspheres were outlined in several papers. It was shown that the main
factor preventing the obtaining and preservation of high Q-factor is
surface atmospheric water adsorption \cite{water}. However, values as high as
$Q\simeq8\times10^{9}$ -- very close to fundamental limit of internal
losses were obtained at He-Ne and near IR region \cite{water,Kimbles}.
Attempts to obtain higher Q moving to minimum of losses of fused silica
$1.55\mu m$, were as for now unsuccessful. Besides chemosorbed water the
reason may be surface scattering on inhomogeneities.

Problems of scattering inside the microsphere and on its surface were
analyzed formerly incompletely. Scattering leads not only to the limitation
of the Q-factor but also lifts degeneracy between degenerate sine and cosine
modes. This effect may be observed as mode doublets \cite{nonlin,lefevre}.
Especially badly explored are questions of surface scattering. In different
papers one may find expressions leading not only to different numerical
estimates but to different functional dependencies from the size of the
resonator and wavelength \cite{first,Kimbles,water,Datsyuk}.

\section{Scattering on internal thermodynamical inhomogeneities
and quality-factor of microspheres}

Intrinsic scattering and absorption losses in microresonators were estimated
previously from the bulk losses as:
\begin{equation}
Q=\frac{2\pi n}{\alpha \lambda},
\label{Qatten}
\end{equation}
where $n$ is index of refraction $\alpha$ is intensity attenuation coefficient
and $\lambda$ is wavelength. However, this approach is not quite accurate for
scattering. We remind here the method of derivation of scattering coefficient
$\alpha$ (see for example \cite{Lines}) to see what modifications should be
made in Eq. (\ref{Qatten}) to take into account specific features of the
microspheres.

Let us divide the whole volume of medium in small volume $d v$,
each having due to fluctuations dielectric constant 
$\epsilon(\vec r)=\delta\epsilon(\vec r)+\epsilon^0$.
Internal small inhomogeneities in the field of the mode behave as dipoles
reradiating light in all direction according to the Rayleigh formula:
\begin{equation}
\frac{I_s}{I}=\frac{\pi^2\sin^2\vartheta}{\lambda^4r^2}
\langle\int\int\delta\epsilon(\vec r_1)\delta\epsilon(\vec r_2)dv_1 dv_2\rangle,
\label{IS}
\end{equation}
where $\vartheta$ is the angle between the dipole axis (coinciding with
polarization of field) and direction of scattering, $r$ is the distance from
scatterer. 

The next step is to integrate Eq.(\ref{IS}) over all angles on large sphere
($r\to\infty$) to obtain total power of scattering. 
\begin{equation}
P_s=I\frac{8\pi^2}{3\lambda^4}
\langle\int\int\delta\epsilon(\vec r_1)\delta\epsilon(\vec r_2)dv_1 dv_2\rangle,
\end{equation}
However, for the microsphere this is not correct. We should take into account
total internal reflection (TIR).

Beams falling on the surface under the angle larger than critical one
$\gamma_0=\arcsin (1/n)$ will either go back in the mode if this angle
lies inside the mode's caustic or will be suppressed in destructive
interference during several reflections. These beams may also go to another
mode, which leads to internal mode coupling, but due to the rareness of
mode spectrum this effect is negligible (specific case of coupling between
oppositely circulating degenerate modes is analyzed below separately). 
In this way, only beams falling under angles less than critical should 
be added to losses. We may not take into account here Frensel transmission
coefficients as these beams may leave the resonator during several reflections.
Conditions for the cutting of angles for $TE$ and $TM$ modes are the following:
\begin{equation}
\sin^2 \gamma_{TE}= \left(\frac{a-d}{a}\right)^2(1-\sin^2\vartheta\cos^2\varphi)<
\frac{1}{n^2}\hspace{20mm}
\sin^2 \gamma_{TM}= \left(\frac{a-d}{a}\right)^2\sin^2\vartheta<\frac{1}{n^2},
\label{cut}
\end{equation}
here $d$ is the distance of the dipole from the surface and $a$ is the radius
of the microsphere. If $d\ll a$, that is always correct for high-Q
whispering gallery modes, the first terms in (\ref{cut}) may be
omitted and hence the result will not depend from the size of the resonator. 
We skip here the following
derivation of $\alpha$ by calculating thermodynamical calculations
as conditions of angle cutting for large microspheres do not interfere with
it (see \cite{Lines}). 
\begin{equation}
\alpha_{is}=\frac{8\pi^3}{3\lambda^4} n^8 p^2 \kappa T\beta_T,
\end{equation}
where $\kappa$ is the Boltzman constant, $T$ is the effective temperature of
glassification ($\sim 1500K$ for fused silica), $\beta_T$ is isothermic
compressibility, and $p$ is Pokkels coefficient of optoelasticity at this
temperature. 

Cutting conditions in this way may be taken into
account by introducing suppression coefficients:
\begin{equation}
Q_{is}=K_{TE,TM}\frac{2\pi n}{\alpha_{is}\lambda}
\label{64jj}
\end{equation}
This coefficient $K_{TE,TM}$ is equal to the relation of complete scattered
power to the power scattered on angles  satisfying conditions
(\ref{cut}).

Numerical calculations for fused silica with $n=1.45$ give
\begin{equation}
K_{TE}=2.8 \hspace{20mm} K_{TM}=9.6
\end{equation}
It follows from these values that TM-modes are less sensitive to intrinsic
scattering losses, but these modes have stronger field on the surface,
and therefore more sensitive to surface inhomogeneities and absorption
on surface contaminations.

\section{Scattering on surface roughness}

To analyze surface scattering we analogously to the previous section
calculate the value of $\alpha_{ss}$, describing losses of travelling
wave per unit length. We start with the same expression in integral form
(\ref{IS}),
but now shall take into account only surface inhomogeneities. As before we should
integrate this expression over angles with account of TIR, but for surface
dipoles the part of light scattered above the surface may go free.
In this way, suppression coefficient may be taken as
$2K_{TE,TM}/(K_{TE,TM}+1)$. In calculations of attenuation coefficient
we as above for generality are not taking it into account and insert only in
final formula for the quality-factor.

Let the wave with intensity distribution $I(y,z)$ travels along a guiding
surface along the local $x$-axis, $y$-axis is chosen also along, and $z$-axis
orthoganally to the surface. Small surface roughness leads to inhomogeneity
of dielectric constant:
\begin{equation}
\delta\epsilon(x,y,z)=(\epsilon_0-1)f(x,y)\delta(z),
\label{vareps}
\end{equation}
here $\delta(z)$ is delta function.
If surface inhomogeneities are weakly correlated and their correlation
function quickly abates to zero on the scale much smaller than the wavelength,
roughness may be described by only two parameters -- variance
$\sigma=\sqrt{<f(x,y)^2>}$ and correlation length $B$. In this case
\begin{equation}
P_s = dx \int I(y,0)
\frac{16\pi^2}{3\lambda^4}
(n^2-1)\pi B^2 \sigma^2 dx=P\alpha dx,
\end{equation}
Thus, considering that the power of the wave is equal to  $P=\int I(y,z)dydz$,
and considering that the wave travels close to the surface, we obtain:
\begin{equation}
\alpha_{ss} = \frac{I(y,0)}{\int I(y,z) dz}
\frac{16(n^2-1)\pi^3 B^2 \sigma^2}{3\lambda^4},
\end{equation}

Now we turn back to microsphere to calculate the ratio of intensities in 
above expression. For simplicity we limit ourselves only with TE-mode, 
which usually have higher quality factor. As the intensity is proportional
to the square of electric field, and fields are described by Bessel
functions, using the same approximations as in \cite{coupl} we obtain for TE-mode:
\begin{equation}
\frac{\int\limits_0^a {{\rm j}^2_\ell(knr)r^2dr}}{a^2 {\rm
j}^2_\ell(kna)} \simeq {a\over2 {\rm j}^2_\ell(kna)}\left(\partial{\rm
j}_\ell(\rho)\over\partial\rho\right)^2_{\rho=kna}\simeq 
{a (n^2-1)\over 2 n^2},
\label{volumetric}
\end{equation}
Finally we obtain expression for the quality factor:
\begin{equation}
Q_{ss}=\frac{K_{TE}}{1+K_{TE}}\frac{3\lambda^3 a}{8n\pi^2B^2\sigma^2}
\end{equation}
We may note that this expression is different from expression obtained in
\cite{Kimbles} but the reason is that the authors underestimated 
volumetric ratio \ref{volumetric} considering it proportional to
$\sqrt{a\lambda}$ and not $a$. In the same paper \cite{Kimbles} authors report on 
measurement of surface roughness of fused silica by means of scanning tunneling
microscope. Values $B=5$nm and $\sigma=1.7$nm were estimated. 
On Fig.1 all calculated limitations on quality factor are plotted, on this
graphics UV and IR absorption in fused silica from literature
\cite{Tosco,Lines} are also taken into account. Bulk losses in fused silica 
were taken as:
\begin{equation}
\alpha \simeq (0.7\mu m^4/\lambda^4+1.1\times10^{-3}\exp(4.6\mu m/\lambda)
+4\times10^{12}\exp(-56\mu m/\lambda)) dB/km
\end{equation}
\begin{figure}
\centerline{\epsfbox{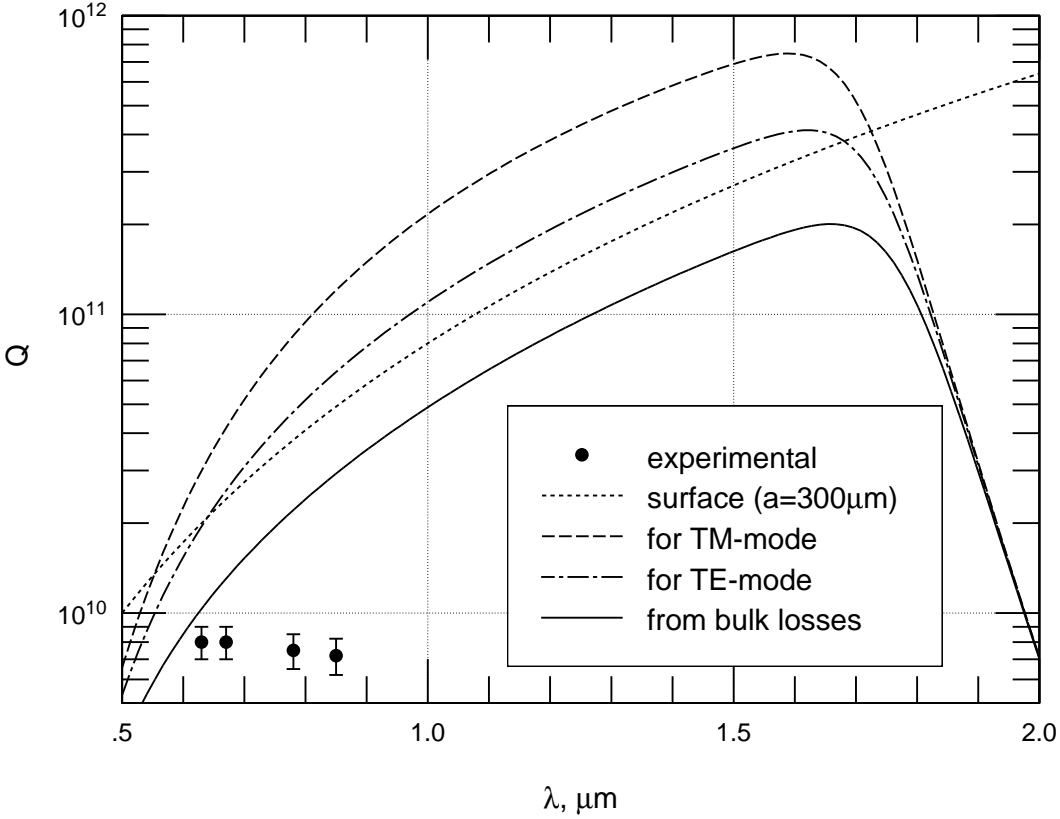}}
\caption
{Influence of different types of losses on the quality factor
of microspere}
\end{figure}
For the estimates of surface scattering resonator radius $a=300\mu m$ was
taken. One can see that for the measured $Q$ in He-Ne range suppression of 
scattering is compensated by surface scattering. For longer wavelengthes, 
it seems, additional experiments are required. For very large spheres (several
millimiters in diameter), quality factor sufficiently higher then $10^{11}$
may be obtained. It seams that even $Q\simeq10^{12}$ is not impossible as
Rayleigh scattering may be lowered by heat treatment on 25\% \cite{Sakaguchi}. 
The only unknown factor is surface absorption on hemosorbed layers
of $OH^-$ ions and water. It should scale linearly with $a$ as surface
scattering (estimate in \cite{Kimbles} is also based on the wrong volumetric
ratio and in this way incorrect) but its dependence on $\lambda$ is unknown.
It is known, however, that in the bulk hydroxil ions lead to vibronic
absorption peak at $\lambda=2.73\mu m$ and at obertones at 1.37, 0.95, 0.725 
and $0.586\mu m$.

Nevertheless the problem of surface scattering in microsperes is not yet
closed neither theoretically nor experimentally.
It is reasonable to suggest that surface roughness may be attributed to
surface capillary waves frozen during solidification. These waves leads to
fluctuations:
\begin{equation}
f(\theta,\phi)=\sum_{L,M}b_l Y^M_L(\theta,\phi),
\end{equation}
where $Y_L^M$ is spherical angular function, and $L>1$.
If according to thermodynamics each wave has energy $\kappa T$, then
\cite{Datsyuk}
\begin{equation}
<b_L^2>=\frac{\kappa T}{\tilde\sigma(L-1)(L+2)},
\end{equation}
where $\tilde\sigma$ is the coefficient of surface tension $\sim 200$ dyn/cm
for fused silica at temperature $T=1500K$.
Though as estimates show the size of these fluctuations will be several
times less than measured in \cite{Kimbles}, correlation function, calculated
for such inhomogeneities has logarithmic shape and in this way may not be
characterized by correlation length. Fluctuations on the scale of wavelength
are of the same order as in nanoscale. In other words, our approach will not
work for this case. Unfortunately the estimate done in \cite{Datsyuk} for
scattering on capillary waves $Q_{cs}\simeq {a\lambda}/((n^2-1)b_L^2)$ is
also incorrect (it was shown \cite{perturb} and experimentally confirmed
\cite{strain} that perturbation with M=0 (ellipticity if L=2), do not
perturb in the first order of approximation quality factor at all).
The problem of frozen capillary waves in any case deserves special
consideration.

\section{Coupled modes in microspheres}
Mode coupling in microspheres due to surface and internal inhomogeneities may
be described using variational approach.
Random deviations of dielectric constant may be written in the form:
\begin{equation}
\delta\epsilon=f(\theta, \phi) F(r),
\label{2t}
\end{equation}
where $F(r)$ -- is random radial function, and $f(\theta, \phi)$ -- is random
angular function.
In particular case of small surface roughness when random fluctuations of the
surface of the sphere may be described as:
\begin{equation}
r(\theta, \phi)=a+f(\theta, \phi),
\label{4t}
\end{equation}
and expression (\ref{2t}) may be written as
\begin{equation}
\delta\epsilon=(n^2-1)f(\theta, \phi) \delta(r-a).
\label{3t}
\end{equation}

From the Maxwell equation, wave equation for the fields inside the microsphere with inhomogeneities may
be obtained:
\begin{equation}
\Delta E-\left(\frac{\epsilon^0(\vec r)}{c^2}+
\frac{\delta\epsilon(\vec r)}{c^2}\right )\frac{\partial^2\vec E}{\partial t^2}
= 0
\label{5t}
\end{equation}

The solutions of unperturbed equation without inhomogeneities
(if $\delta\epsilon=0$) have the form
\begin{equation}
\vec E_j=e^(-i\omega_j t)\vec e(r,\theta,\phi),
\label{6t}
\end{equation}
where $\vec e_j(r,\theta, \phi)$ -- is vector harmonic,
satisfying Helmholtz equation:
\begin{equation}
\Delta\vec e_j+\epsilon^0 k_j^2\vec e_j=0
\end{equation}
and index $j$ corresponds to all possible types of oscillations, and $j=0$
corresponds to initially excited mode.
Using the method of slowly varying amplitudes we find solution as:
\begin{equation}
\vec E= e^{-i\omega_0 t} \sum A_j(t) \vec e_j
\end{equation}
After substituting this sum in equation (\ref{5t}) and omitting small terms
we obtain:
\begin{equation}
2i\omega_0\epsilon^0\sum \frac{dA_j(t)}{dt}\vec e_j+
\omega_0^2 \delta\epsilon \sum A_j(t)\vec e_j
+ \epsilon^0\sum_j (\omega_j^2-\omega_0^2) A_j(t)\vec e_j = 0
\end{equation}
After multiplication of this equation on $\vec {e_j}^*$ and integration over
the whole volume, with account of modes' orthoganality, we obtain usual
equations for coupled modes:
\begin{equation}
\frac{dA_k}{dt}+i\Delta\omega_k A_k=i\sum_j A_j\beta_{jk},
\end{equation}
where $\Delta\omega_{k}=\omega_k-\omega_0$ and
\begin{equation}
\beta_{jk}=\frac{\omega_0}{2n^2}\frac{\int \vec {e_j} \delta\epsilon \vec{e_k}^* dv}
{\int |\vec {e_j}|^2 dv}
\label{beta}
\end{equation}
In this expression it is the random function $\delta\epsilon$ which leads to
the coupling between $\vec e_j$ and $\vec e_k$. We are interested only
on the modulus of the coefficient of $\beta_{jk}$, which determines the rate
of energy redistribution between different modes.
If the value of inhomogeneities and their correlation length are very small if
compared with wavelength we may average $\beta^2_{jk}$ and obtain that:
\begin{equation}
\beta_{jk}^2=\frac{\omega^2_0}{4n^4}\frac{\langle\int\delta\epsilon(\vec r)\delta\epsilon(0)dv\rangle}{V_{jk}},
\label{beta3}
\end{equation}
where $V_{jk}$ -- is overlap volume of modes:
\begin{equation}
V_{jk}=\frac{\int|\vec {e_j}|^2 dv \int|\vec {e_k}|^2 dv}{\int|\vec {e_j}|^2 |\vec {e_k}|^2 dv}.
\label{beta3a}
\end{equation}
In the most interesting case of coupling between two modes $A_+(t)$ and
$A_-(t)$, travelling inside the microsphere in opposite direction, fields'
distribution for these two modes differs only on phase factor
$\exp(\pm i m\phi)$. In this way $\vec e_j = {\vec e_k}^*$ and $V_{jk}$ in
this case transforms in effective volume of field localization:
\begin{equation}
V_{eff}=\frac{(\int|\vec {e_j}|^2 dv)^2}{\int|\vec {e_j}|^4 dv},
\label{beta4}
\end{equation}

\section{Effects of coupled modes}

To analyze the consequences of internal coupling between modes travelling in
opposite directions on the output characteristics of a resonator we may
use the same quasigeometrical approach that we used in \cite{coupl} to
analyze coupler devices for the microspheres. For simplicity we analyze here
only the case of ideally mode matched (or monomode) coupler device.
The set of equations for the internal and external amplitudes looks as
follows:
\begin{eqnarray}
\frac{{\rm d } A_+}{{\rm d}t}&+&(\delta_0+\delta_c+i\Delta\omega)A_+=
iA_-\beta+i\frac{T}{\tau_0} B_{in} \nonumber\\
\frac{{\rm d } A_-}{{\rm d}t}&+&(\delta_0+\delta_c+i\Delta\omega)A_-=
iA_+\beta \nonumber\\
B_t&=&\sqrt{1-T^2}B_{in}+iTA_+\nonumber\\
B_r&=&iTA_-
\label{difset}
\end{eqnarray}
\begin{figure}
\centerline{\epsfbox{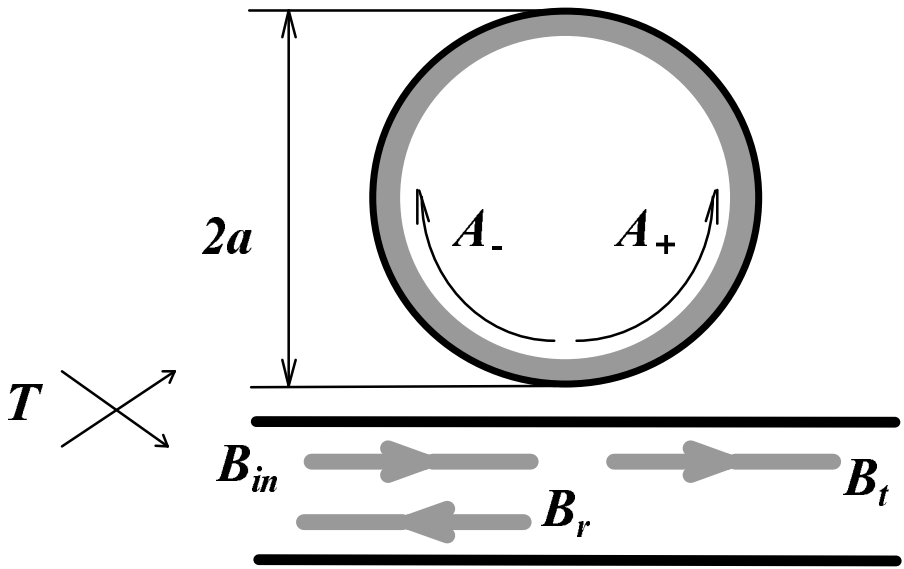}}
\caption
{Backscattering in a microspere}
\end{figure}
Where $A_+(t)$ and $A_-(t)$ are as before the amplitudes of oppositely circulating
modes of TIR in the resonator (Fig. 2) to model the
whispering gallery modes. $B_{in}$ is the amplitude of pump and $B_t$ and
$B_r$ are output amplitudes transmitted and reflected in coupler. $T$ is
the amplitude transmittance coefficient, describing coupler,
$\delta_0=2\pi n/\alpha \lambda$ is the decrement of internal losses,
$\delta_c=T^2/2\tau_0$ is the decrement of coupler device, $\tau_0$ is the
circulation time $\tau_0\simeq 2\pi n a/c$, and $\Delta\omega$ is frequency
detuning from unperturbed resonance frequency $\omega_0$ (for details see
\cite{coupl}).
The stationary solution of Eq.(\ref{difset}) is the following
\begin{eqnarray}
A_+&=&\frac{i}{T}\frac{2\delta_c\beta}{(\delta_0+\delta_c)^2+\beta^2-\Delta\omega^2+i2\Delta\omega(\delta_0+\delta_c)}B_{in}\nonumber\\
A_-&=&-\frac{1}{T}\frac{2\delta_c(\delta_0+\delta_c+i\Delta\omega)}{(\delta_0+\delta_c)^2+\beta^2-\Delta\omega^2+i2\Delta\omega(\delta_0+\delta_c)}B_{in}\nonumber\\
B_t&=&\frac{\delta_0^2-\delta_c^2+\beta^2-\Delta\omega^2+i2\delta_0\Delta\omega}{(\delta_0+\delta_c)^2+\beta^2-\Delta\omega^2+i2\Delta\omega(\delta_0+\delta_c)}B_{in}\nonumber\\
B_r&=&-\frac{i2\delta_c\beta}{(\delta_0+\delta_c)^2+\beta^2-\Delta\omega^2+i2\Delta\omega(\delta_0+\delta_c)}B_{in}
\label{couplampl}
\end{eqnarray}
If internal mode coupling constant is weaker than attenuation
$\beta<\delta_0+\delta_c$ than Eq.\ref{couplampl} has only one resonance at
$\Delta\omega=0$ and backscattering is small. The situation is not very
different in this case from the case of one mode analyzed in \cite{coupl}.
In temporal language it means that internal coupling simply has no time to
build backscattered wave during the ringdown time. Interesting is, however,
that the regime of critical coupling (when $B_t=0$) is shifted and obtained
not for $\delta_c=\delta_0$ but for $\delta^2_c=\delta_0^2+\beta^2$ and in
this case not all input power is lost in the resonator but some part of it
reflects back in the coupler.
\begin{equation}
B_r=i\frac{\beta}{\delta_0+\delta_c}
\end{equation}
This means that ringdown time is equal to the time
needed to repump circulating mode in oppositely circulating one.
\begin{figure}
\centerline{\epsfbox{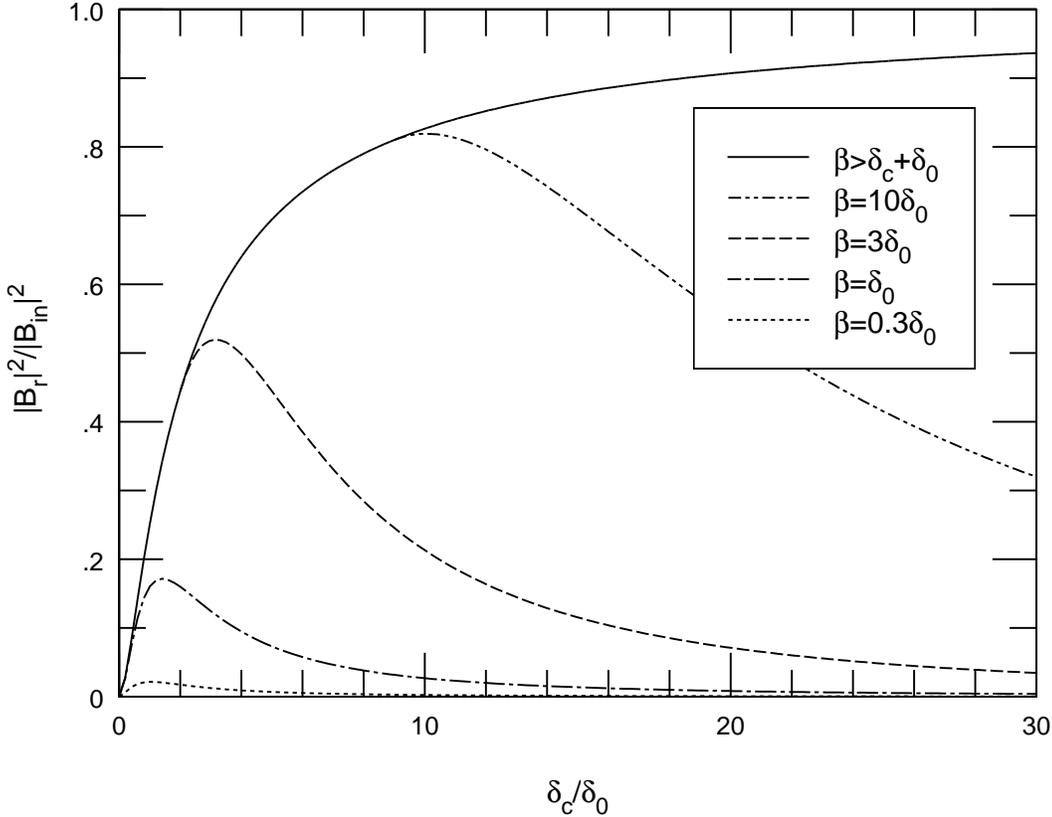}}
\caption
{The dependence of power, reflected in coupler due to backscattering in
microsphere, from loading}
\end{figure}
\begin{eqnarray}
|A_+|^2&=&|A_+|^2=\frac{1}{T^2}\frac{\delta_c^2}{(\delta_0+\delta_c)^2}B^2_{in}\nonumber\\
|B_t|^2&=&\frac{\delta_0^2}{(\delta_c+\delta_0)^2}B_{in}\hspace{20mm}
|B_r|^2=\frac{\delta_c^2}{(\delta_c+\delta_0)^2}B_{in}\nonumber\\
\label{couplmagn}
\end{eqnarray}
The case of $\beta\geq\delta_0+\delta_c$ is much more interesting and even
leads to somewhat unexpected results. In this case there are two resonances
at frequencies $\Delta\omega=\pm\sqrt{(\beta^2-(\delta_0+\delta_c)}$ i.e.
internal coupling lifts degeneracy between sine an cosine standing modes in the
microsphere and there is enough time to form them. All intensities in resonances
in this case do not depend on $\beta$:

What is interesting and inevident that if $\delta_0\ll\delta_c$ (overcoupling)
but still $\delta_0+\delta_c<\beta$ then most part of input power is
backscattered and transmitted power tends to zero. This property may become
extremely valuable for future applications of microspheres in laser stabilization.
To verify this result additional experiments are required, but we saw many times that
backscattering is practically absent when there is no splitting and practically
does not depend on loading when doublets are clearly seen (see Fig.3).

\section{Calculation of mode splitting on internal and surface inhomogeneities}
Mode coupling leads to splitting of initially degenerate modes, if
$\beta$ constant is much larger than mode decrement of internal and
coupling attenuation $\delta_0+\delta_c$ then:
\begin{equation}
\frac{\Delta\omega}{\omega}=\frac{2\beta}{\omega_0},
\label{beta4a}
\end{equation}
If thermodynamical inhomogeneities are calculated in the same way as before:
\begin{equation}
\left(\frac{\Delta\omega}{\omega}\right)_{is}=
\sqrt\frac{n^4 p^2 \kappa T\beta_T}{V_{eff}}=
\sqrt\frac{3\lambda^4\alpha_{is}}{8\pi^3 n^4 V_{eff}},
\label{beta5}
\end{equation}
in agreement with qualitative estimates in \cite{haroche}.
Effective volume of the most interesting $TE_{\ell \ell 1}$-mode may be
calculated according to the formula \cite{first}
\begin{equation}
V_{eff}=2.3 n^{-7/6}a^{11/6}\lambda^{7/6},
\label{beta6}
\end{equation}
In this way, for $TE_{\ell\ell1}$-mode in fused silica microsphere:
\begin{equation}
\left(\frac{\Delta\omega}{\omega}\right)_{is}\simeq \frac{5\times10^{-7}\mu m^{3/2}}{\lambda^{11/12}a^{7/12}},
\label{isplit}
\end{equation}
If $\ell\neq m$ the following asymptotic approximation is valid:
\begin{equation}
V_{eff,\ell m}=V_{eff,\ell\ell}(1+0.5\sqrt{\ell-m-0.5}),
\label{beta8}
\end{equation}

Now let us analyze the case of mode splitting due to the surface
inhomogeneities. From (\ref{beta}) and (\ref{vareps}) after averaging
\begin{equation}
\beta_{ss}^2=
\frac{\omega^2_0}{4n^4}\frac{\pi B^2 \sigma^2 |\vec e|^4}
{V_{eff}\int |\vec e(r)|^4 dr}
\label{beta9}
\end{equation}

or for $TE_{\ell\ell1}$-mode in fused silica microsphere:
\begin{equation}
\left(\frac{\Delta\omega}{\omega}\right)_{ss}\simeq \frac{1.1\sigma B}{\lambda^{1/4}a^{7/4}},
\label{ssplit}
\end{equation}
It is easy to show that for measured size of surface inhomogeneities
\cite{Kimbles} this expressions gives substantially lower level of
coupling between modes than internal inhomogeneities (\ref{isplit}).
On (Fig. 3) results of calculation for fused silica microsphere for
$\lambda=0.63 \mu m$ according to (\ref{isplit},\ref{ssplit}) are shown.
\begin{figure}
\centerline{\epsfbox{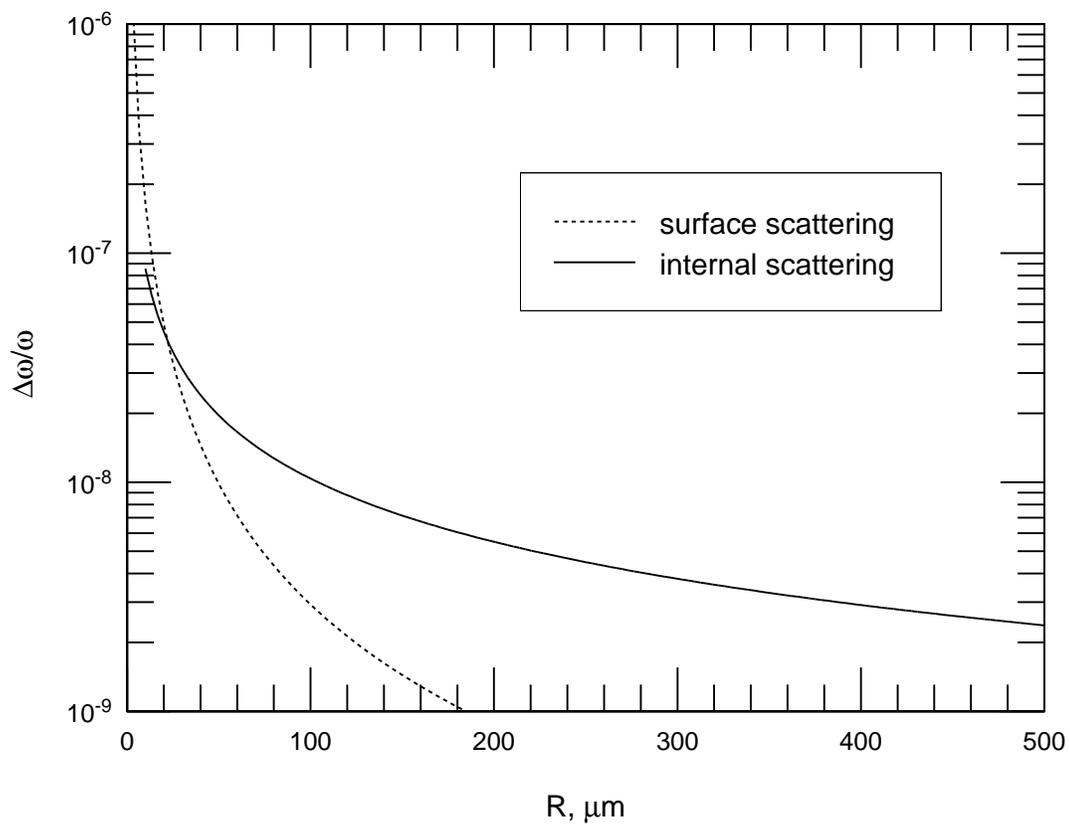}}
\caption
{Normal mode splitting on inhomogeneities in microsperes}
\end{figure}

\acknowledgments
This research was supported in part by Russian Foundation for
Basic Research (grant \#96-15-96780).


\end{document}